\documentclass[preprint]{aastex}
\usepackage{graphicx}

\def\et{{\it et al.}}

\def\cp{{\rm CAPMAP }}

\begin{document}

\date{8 pages, 7 figures. To be published in the proceedings of ``The Cosmic Microwave Background and its Polarization'', New Astronomy Reviews (eds. S. Hanany and K. A.Olive).}

\title{{\bf The CAPMAP Instrument and its First Season}}

\author{Denis Barkats\footnote{On behalf of the \cp~collaboration. Email: dbarkats@princeton.edu}}


\begin{abstract}
I describe here the new \cp~({\bf C}osmic {\bf A}nisotropy {\bf P}olarization {\bf MAP}er) instrument, which performed its first season of observing between February and April 2003 from the Crawford Hill 7-meter antenna in NJ. \cp is based on the design for the PIQUE instrument, but has better sensitivity and superior angular resolution.

\end{abstract}

\section{INTRODUCTION}

 The status of measurements of CMB polarization has evolved rapidly. Only a year separates the publication of tight upper limits~\cite{pique01,pique02,polar} and the detection  of polarization ~\cite{dasi02}. In the summer of 2001, we started designing the \cp instrument. \cp is a 16-element array comprised of twelve 90~GHz and four 40~GHz correlation receivers (four 90~GHz elements have been deployed for initial observations since February 2003). A new feed system has been designed to couple \cp to the 7-meter Crawford Hill antenna to  produce a 4$^\prime$ beam. The goal is to concentrate sensitivity on a  $1^\circ$ diameter cap around the North celestial pole (NCP). As indicated in Figure~\ref{fig:senfac}, \cp will probe higher angular scales than those probed by MAP and DASI, and will be sensitive to the predicted peak in polarization.
 I will discuss in more details specifics of the telescope, the optics, and the receivers.

\
\section{OPTICS}
\subsection{Telescope}
\begin{figure}[!h]
\begin{center}
  \includegraphics[scale=0.65]{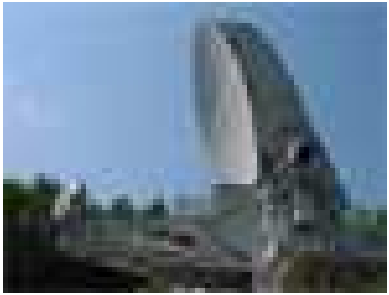}
  \caption{\footnotesize Photograph of the 7-meter Crawford Hill antenna pointing at the horizon. It is located at a Lucent Technologies site in Holmdel, NJ (lat$=40.392^\circ$N, lon$=74.186^\circ$W and 119 meters of elevation.) }\label{fig:tele}
\end{center}
\end{figure}
In order to achieve  high angular resolution, \cp uses the Lucent Technologies 7-meter offset Cassegrain antenna located in Holmdel, NJ at the Crawford Hill Observatory. Built in 1977, it was originally designed to test polarization multiplexing of signals to geosynchronous satellites at K band and for radio astronomy up to 300 GHz. As a result, it demonstrates many of the requirements necessary for CMB measurements:
\begin{itemize}
    \item low sidelobe level.
    \item low cross polarization level.
    \item good pointing accuracy and tracking capabilities on an alt-az mount.
    \item  large f/D ratio, large useable focal plane.
    \item temperature-controlled Cassegrain focus room.
\end{itemize}    
A detailed description of the antenna can be found in the Bell Labs Technical Journal~\cite{chu}.
Prior to the summer of 2001, when we started using the antenna, it had primarly been used during the last eight years to train winter-over operators for the south pole AST/RO telescope. We invested large amounts of work in the following two years to repair and upgrade the various mechanical, electrical, and software systems. 

\subsection{Feed system}

\begin{figure}[!h]
\begin{center}
  \includegraphics[scale=0.5]{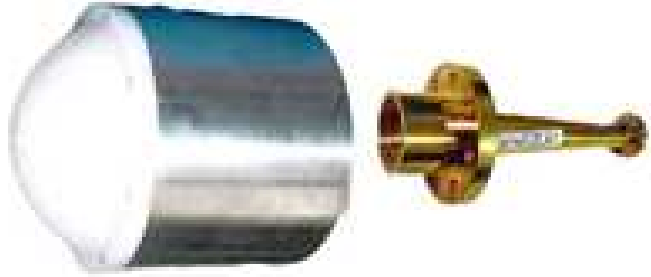}
  \caption{ \footnotesize Photograph of horn and lens. The lens is attached to a metal shroud which in turn attaches  to the horn via 3/4" G-10 spacers. These spacers keep the horn tied to the second stage of the fridge at $\sim$~20K and the lens to the first stage  at $\sim$~100K. }\label{fig:hornandlens}
\end{center}
\end{figure}

If the full aperture were used, the 7-meter antenna would produce a $0.02^\circ$ beam in the sky. This resolution is  smaller  than needed to probe the scales where the polarization peaks, at 10$^\prime$ or $\ell\sim$~1000. Thus, the feed system is designed to produce a  FWHM beam size of 4$^\prime$ and to minimize pickup of stray radiation from the ground. 
This is achieved by aggressively under-illuminating  both mirrors. The secondary and primary edge illuminations are respectively -56dB and -30dB. The small ($\sim 3^\circ$) beam pattern required to produce such low edge tapers is the main price that must be paid for the advantage of the large effective focal ratio of $f/D=5.6$. A standard small-angle corrugated horn  had to be rejected as a feed because its size would make it prohibitive to cryogenically cool and because its sidelobe pattern could not be controlled to the desired level. 

\begin{figure}[!ht]
\begin{center}
  \includegraphics[scale=0.5]{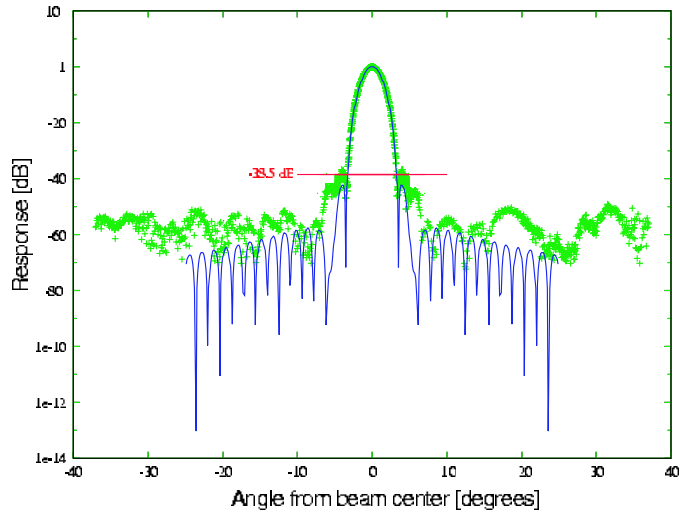}
  \caption{\footnotesize Measured (green points) and predicted (blue line) beam pattern response  versus azimuthal angle for the horn plus lens feed system. The  first sidelobe is 38.5~dB below the main beam. We believe the discrepancy of the beam map from the prediction (especially the far sidelobe on the right) are reflections from walls surrounding the beam map setup. Courtesy of Jeff McMahon.}\label{fig:beammap}
\end{center}
\end{figure}

The solution  for the feed optics is a 3~cm aperture corrugated feedhorn followed by a 12~cm diameter HDPE meniscus lens shown in Figure~\ref{fig:hornandlens}. The lens is held in front of the horn with a metal shroud. The shroud is coated  with a $1/4^{\prime\prime}$ thick microwave absorbing honeycomb to prevent stray radiation from illuminating the horn. This feed system produces the necessary beam size ($2.6^\circ$ FWHM  at 90 GHz) to achieve the designed secondary illumination and was optimized to keep its sidelobes below $\sim$~-40~dB (see Figure~\ref{fig:beammap}).

We note that because this antenna in not surrounded  by ground screens, the extreme under-illumination and the obsessive control of sidelobes are necessary design features to keep the spillover below levels that would interfere with  the observation of CMB polarization.

\subsection{Array configuration}
\begin{figure}[!h]
\plottwo{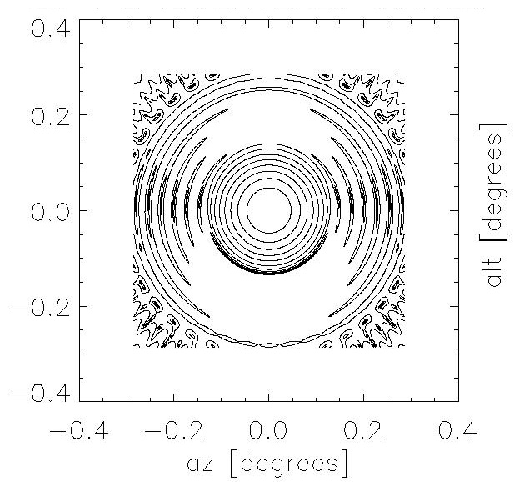}{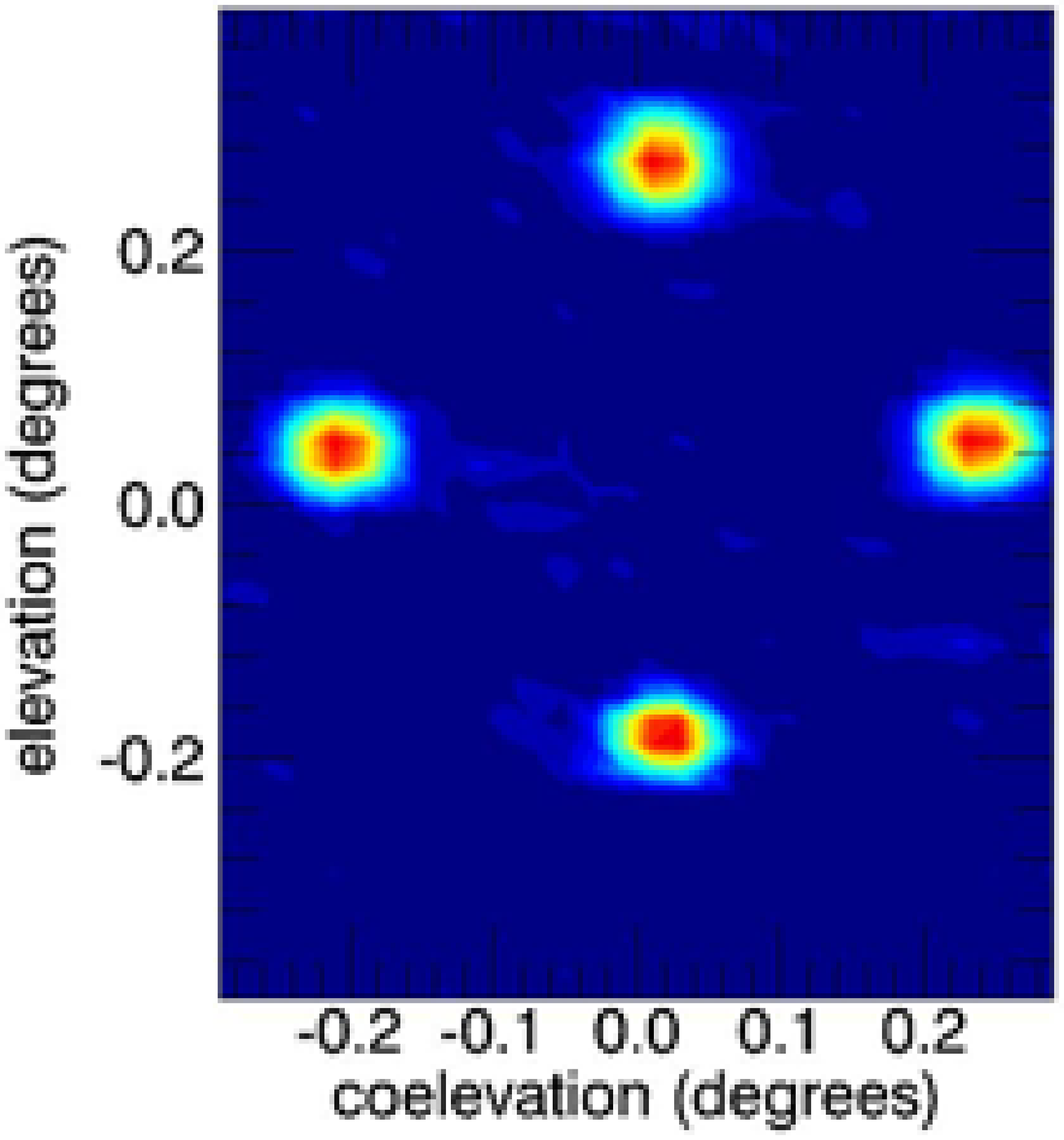}
\epsscale{10}
\caption{ \footnotesize Left: Simulated beam map of a receiver 39 cm below the focal point. No distortions from a symmetric gaussian are visible down to the first sidelobe level. The features in the corners are artifacts of the code. Right: Observed beam map of our array using Jupiter from February 12, 2003. The predicted position of each beam agrees well with the calculated plate scale of 15 cm $\Rightarrow 0.25^\circ$}\label{fig:beam}
\end{figure}

Because of its slow optics, the Crawford Hill telescope is well suited for an array of receivers at its focus. To determine the size of distortions for receivers displaced from the focal point, we performed extensive studies using both measurements from receivers at different positions in the focal plane and physical optics simulations~\cite{grasp8}. Figure~\ref{fig:beam} (left) shows a simulated beam contour plot for a receiver located 39 cm above the focal point and pointed towards the center of the secondary. Distortions remain small even this far from the focal point.

The plate scale for the focal plane is such that a 15 cm shift in the focal plane corresponds to a $0.25^\circ$ shift of the beam in the sky.  For the initial deployment of four radiometers, we arranged the  receivers in a 15 cm radius diamond pattern, obtaining the  composite beam map in Figure~\ref{fig:beam} (right). This configuration allows us to observe a $1^\circ$ diameter cap around the NCP by scanning the telescope back and forth in azimuth. This scan strategy naturally provide many systematic checks.

\section{RECEIVERS}

The receivers are heterodyne correlation polarimeters. With approximately 100K system noise and 14 GHz bandwidth, the receiver sensitivity is $\sim$1~mK$\sqrt{s}$. The signal is amplified by two back-to-back  MMIC LNAs\footnote{Monolithic microwave integrated circuit low noise amplifiers}(see section~\ref{sec:mmic}). After being  down-converted to IF (2-18~GHz), filtered, and further amplified, the signal is  detected with a correlation multiplier.  The signal is phase-switched in the LO at 4~kHz  to avoid the $1/f$ noise of the LNA, which have a $1/f$ knee before phase switching of $\sim$1~kHz. The IF is divided into three sub-bands to allow  frequency discrimination. A single LO signal is generated and then distributed to each receiver after amplification in the MMIC power amplifiers.
\begin{figure}[!h]
\begin{center}
  \includegraphics[scale=0.4]{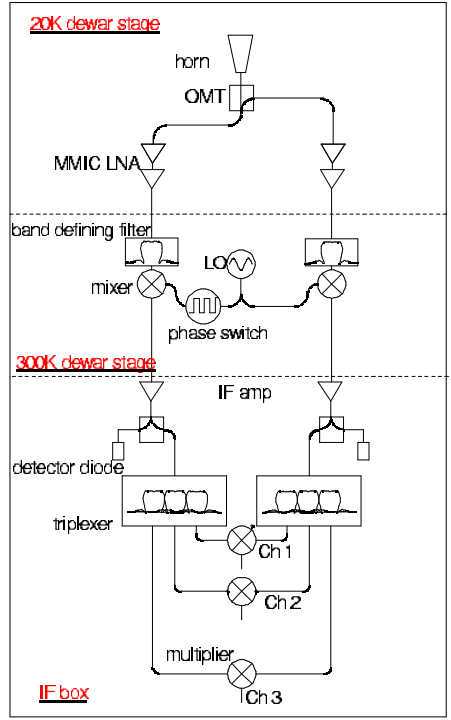}
  \caption{\footnotesize Schematics of a receiver. In the top section cooled to 20K, the signal is split into two orthogonal linear polarizations with an OMT, then amplified in the MMICs. The middle section is inside the dewar. There, the signal is filtered and down-converted before exiting the dewar to the IF box where the two arms are finally recombined in the correlation multiplier.}\label{fig:receiver}
\end{center}
\end{figure}

\subsection{MMIC LNAs}
\label{sec:mmic}
\begin{figure}[!ht]
\epsscale{1}
\plottwo{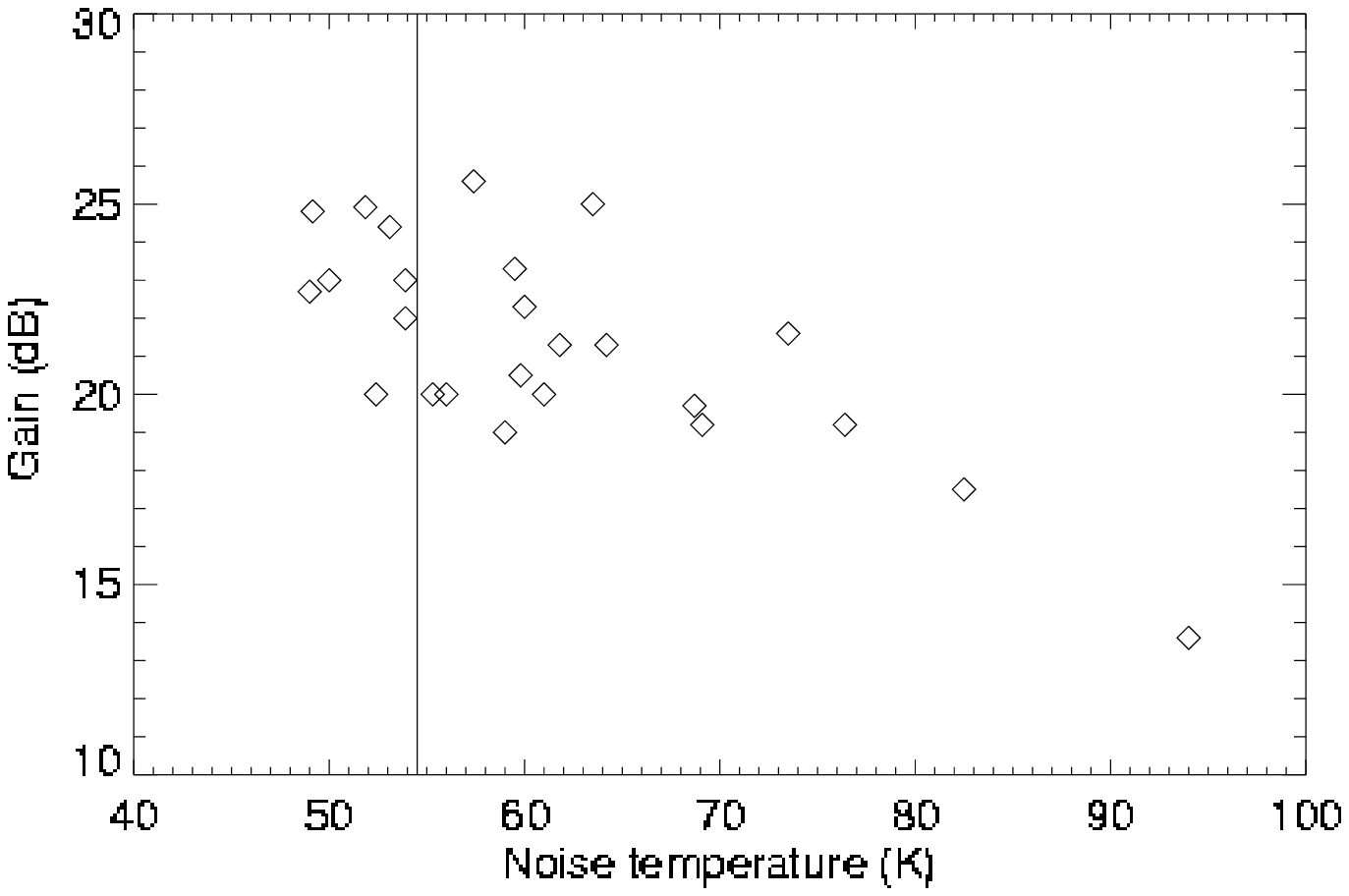}{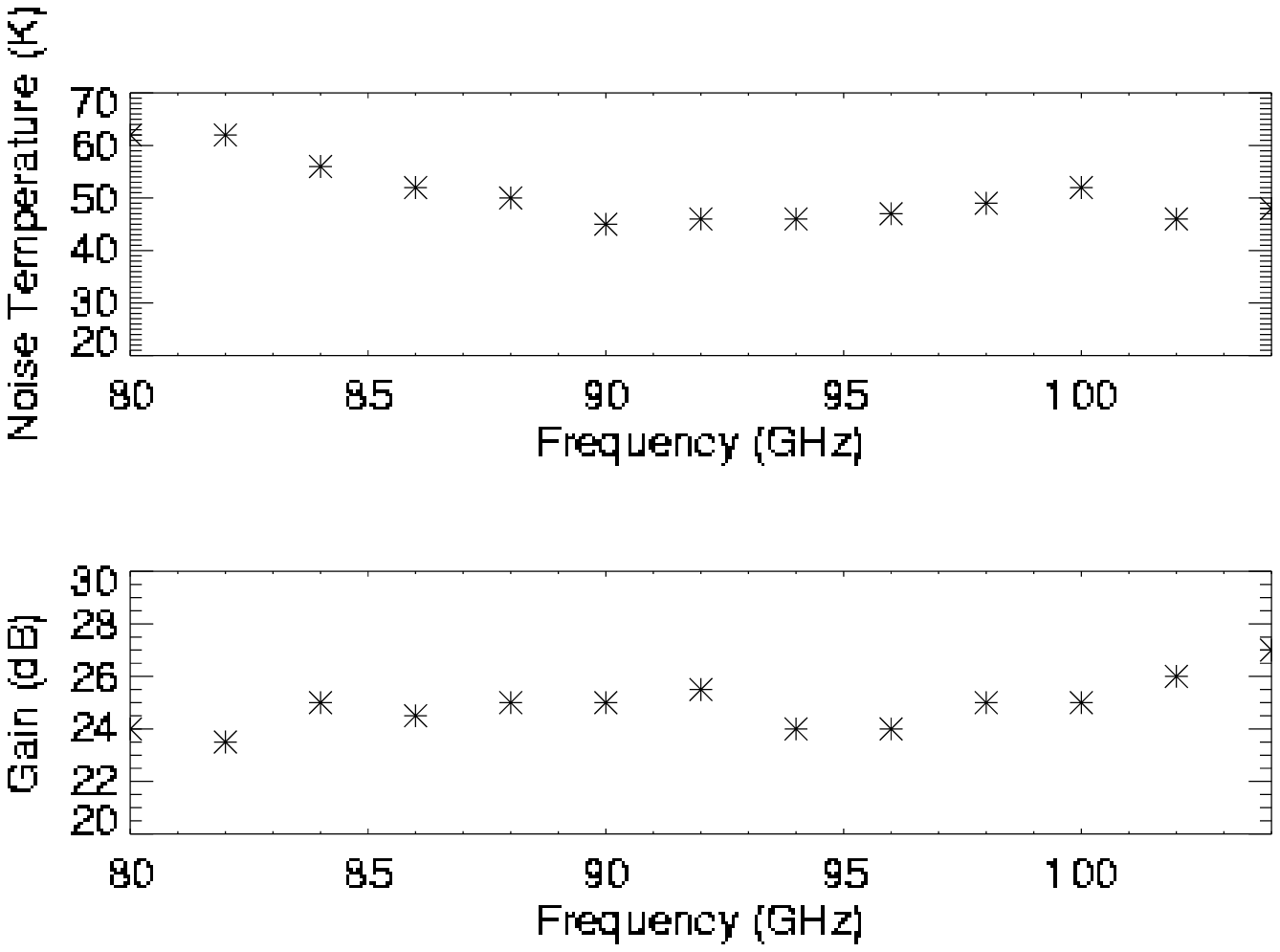}
\caption{ \footnotesize Left: Distribution of noise temperature and gain for the CAPMAP MMICs. Only the best devices (left of the vertical line) were used as front ends. Right: Noise and gain measurement versus frequency of a typical MMIC. }\label{fig:mmicfigs}
\end{figure}

We take advantage of recent microwave technology developments by using a new generation of HEMT amplifier. We use MMIC LNAs and power amplifers  developed, integrated, and tested at JPL (by Todd Gaier). We cascade two MMIC LNAs together to obtain the necessary gain. Figure~\ref{fig:mmicfigs}~(right) shows the noise temperature and gain  versus frequency for a typical MMIC cooled to 20K. Figure~\ref{sec:mmic}~(left) shows the distribution of noise and gains for the devices we have acquired so far. Note that 20dB of gain and less than 55K noise temperature are not uncommon. 

\subsection{Data Acquisition}
The CAPMAP support electronics and data acquisition modules are all housed in the focal point cabin one meter away from the dewar. The vertex cab, support electronics, data acquisition module, and both dewar stages are temperature controlled. All such critical environmental and cryogenic temperatures are monitored and recorded at 1 Hz. All radiometer channels are synchronously sampled at 96 kHz using an ICS-610\footnote{http://www.ics-ltd.com/} data acquisition board. These data are then averaged down to 4 kHz, demodulated in software, and recorded at 100~Hz. For this season, the azimuth scan rate was one beam per 0.3 seconds.

\section{CONCLUSION}

\begin{figure}[!h]
\begin{center}
  \includegraphics[scale=0.5]{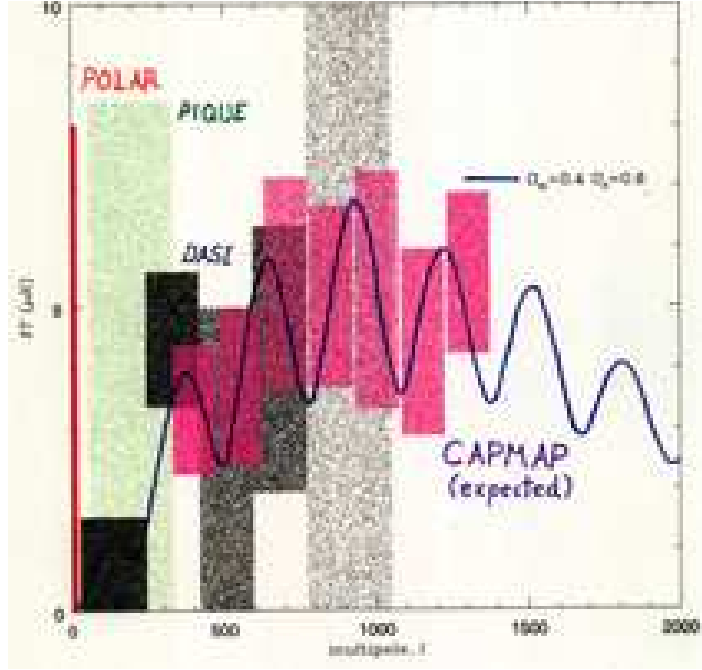}
  \caption{\footnotesize Expected EE polarization power spectra for the full 16 element \cp instrument for two observing seasons. The POLAR (red), PIQUE (green), and DASI (black) results are shown for comparison. Courtesy of Matt Hedman.}\label{fig:senfac}
\end{center}
\end{figure}

CAPMAP complements MAXIPOL, B2K, CBI, DASI, and MAP in  frequency coverage,  experimental technique and angular resolution. An example of  the sensitivity of \cp to CMB polarization anisotropy is given in Figure~\ref{fig:senfac}. 
For this season, four receivers were deployed and observed during fair weather between January 16 and April 30, 2003 with sensitivities near the predicted ones based on 100~K system noise and 14~GHz bandwidth. As indicated in Figure~\ref{fig:beammap}, the beams are symmetric with acceptable sidelobes. Results from this first campaign will be detailed in a forthcoming publication. We expect to deploy the full 16 element array in early November 2003.

\subsection{Acknowledgements}

CAPMAP is a collaborative effort involving researchers at various institutions: Jeff McMahon, Phil Farese, Suzanne Staggs, and myself at Princeton University; Keith Vanderlinde, Colin Bischoff, Matthew Hedman, Dorothea Samtleben, and Bruce Winstein at University of Chicago; Eugenia  Stefanescu and Josh Gundersen at University of Miami; and Todd  Gaier at JPL.

This work was supported by the NSF grants  \#PHY-0099493 , \#PHY-9984440, \#PHY-0114422.

\pagebreak

\end{document}